\documentclass[12pt]{article} 
\usepackage{epsfig,amsmath,euscript}
\usepackage{color}
\begin{document}

\begin{titlepage}

\vspace{0.7cm}
\begin{center}
\Large\bf 
Positive parity pentaquark towers in large $N_c$ QCD
\end{center}

\vspace{0.8cm}
\begin{center}
{\sc Dan Pirjol$^a$ and Carlos Schat$^{b,c}$}\\
\vspace{0.7cm}
{\sl $^a$Center for Theoretical Physics, Massachusetts Institute of
Technology, Cambridge, MA 02139, USA\\[0.3cm]
$^b$CONICET and Departamento de F\'{\i}sica, FCEyN, Universidad de Buenos Aires,
Ciudad Universitaria, Pab.1, (1428) Buenos Aires, Argentina \\[0.3cm]
$^c$Departamento de F\'{\i}sica, Universidad de Murcia, E-30071 Murcia, Spain}
\end{center}

\vspace{1.0cm}
\begin{abstract}
\vspace{0.2cm}\noindent
We construct the complete set of positive parity pentaquarks, which
correspond in the  quark model to  $\bar s q^{N_c+1}$ states with one
unit of orbital angular momentum $L=1$. In the large $N_c$ limit they
fall into the ${\cal K}=1/2$ and ${\cal K}=3/2$ irreps (towers) of the contracted
$SU(4)_c$ symmetry. We derive predictions for the mass spectrum and
the axial couplings of these states at leading order in $1/N_c$.
The strong decay width of the lowest-lying positive parity exotic state is of order $O(1/N_c)$, such that this state is
narrow in the large $N_c$ limit.
Replacing the antiquark with a
heavy antiquark $\bar Q q^{N_c+1}$, the two towers become degenerate,
split only by $O(1/m_{Q})$ hyperfine interactions. 
We obtain predictions for the strong decay widths of heavy pentaquarks
to ordinary baryons and heavy $H^{(*)}_{\bar Q}$ mesons at leading order in 
$1/N_c$ and $1/m_Q$.
\end{abstract}
\vfil

\end{titlepage}

\section{Introduction}

The last few years witnessed a renewed interest in hadronic physics,
originated in part by many new findings in hadron spectroscopy, the most
conspicuous being the narrow pentaquark states reported by more than ten
independent experimental groups~\cite{discovery,posCLAS,SAPHIR,posex,STAR,posexheavy,newposex}.
The narrow state predicted by a chiral soliton model~\cite{DPP97} provided the initial motivation
for the search of the
$\Theta^+$ pentaquark and its possible partners. After the reports of
null results started to accumulate~\cite{nullex,negCLAS,nullexheavy} the initial optimism
declined, and the experimental situation remains ambiguous to the
present. The increase in statistics led to some recent new claims for
positive evidence~\cite{newposex}, while the null
result~\cite{negCLAS} by CLAS is specially significant because it
contradicts their earlier positive result~\cite{posCLAS}, suggesting
that at least in their case the original claim was an artifact due to
low statistics. All this experimental activity spurred a great amount of
theoretical work in all kinds of models for hadrons and a renewed
interest in soliton models. There is a great amount of uncertainty in
model calculations that could be reduced with more
experimental input, like the
spin and parity  of the reported exotic states~\cite{qn} or the possible
existence of spin-flavor partners~\cite{STAR,Capstick:2003iq}.  Lattice QCD calculations
are constantly improving but the situation also remains
ambiguous~\cite{Liu:2005yc,latt32}, in part by the extrapolations to
light masses and the difficulty to disentangle scattering states from
bound states in a finite volume. Given the difficulties still faced by
first principle QCD calculations, the $1/N_c$
expansion~\cite{'tHooft:1973jz} of QCD, where $N_c$ is the number of
colors, provides a very useful analytical tool (for a recent account 
see~\cite{Goity:2005fj}). It is based on the fundamental theory of the
strong interactions, and relates the chiral soliton model to the
more intuitive quark model
picture~\cite{Manohar:1984ys,Jenkins:2004tm,Kopeliovich:2006na,JW},
where the pentaquark correspond to states with quark content $\bar
qq^4$.

In the large $N_c$ limit, QCD has a contracted spin-flavor
symmetry $SU(2F)_c$~\cite{Largenspinflavor,DJM} in the baryon sector, also known as ${\cal K}$-symmetry, 
that constrains their mass spectrum and couplings.
The breaking of the spin-flavor
symmetry can be studied systematically in an $1/N_c$ expansion.
This approach has been applied to the ground-state $[\mathbf{56},0^+]$
baryons \cite{DJM,largenrefs,largen,heavybaryons}, and to their orbital 
excitations~\cite{PY1,Goity,SU3,symml2,Cohen:2003jk,PiSc,Matagne:2004pm}. 
The large $N_c$ expansion has also been applied to hybrid baryons \cite{CPY99} and more recently to  
exotic baryons containing
both quarks and antiquarks~\cite{Jenkins:2004vb,CoLe,MW,PiSc2}. 

In this paper we assume the existence of these exotic states, and investigate
their properties in the case that they have positive parity. A partial subset
of these states were considered in the $1/N_c$ expansion in Ref.~\cite{Jenkins:2004vb}.
Negative parity exotic states have been studied in~\cite{StWi,MW,PiSc2}.
A brief report of some results presented here has been given in~\cite{Pirjol:2006hg}.
We start by constructing the color singlet $\bar s q^{N_c+1}$ states by
coupling the spin-flavor, orbital and color degrees of freedom, all constrained by Fermi statistics.
The light exotic states we obtain are members of the ${\cal K}=\frac{1}{2}$ 
and ${\cal K}=\frac{3}{2}$ irreducible representations of the contracted spin-flavor 
symmetry. This extends the analysis of \cite{Jenkins:2004vb}, which considered only the
first irreducible representation.

Similar states with one heavy antiquark exist, that can be labelled in the large $N_c$
limit by the conserved quantum number ${\cal K}_\ell$ associated with the light degrees of
freedom. In our case of the positive parity pentaquarks, there is only one tower with
${\cal K}_\ell = 1$, containing all nonstrange states for which the isospin $I$ and spin of the
light degrees of freedom $J_\ell$ satisfy $|I-J_\ell | \leq {\cal K}_\ell \leq I+J_\ell $.
An important difference with respect to the treatment of the strong decay
amplitudes done in Ref.~\cite{Jenkins:2004vb} is that we will keep the orbital 
angular momentum explicit in the transition operator, which is required to get the
correct $N_c$ scaling of the relevant couplings.

Although the existence of pentaquarks is not yet established or completely 
ruled out by experiments, one thing that seems to be fairly well established 
is that if they exist they should be narrow, with a width of $\Gamma \le 1~{\rm MeV}$, otherwise
they would have been observed long 
before~\cite{bounds}. Explanations for the uncanny narrow width vary. 
Cancellations between coupling constants have been invoked in the context of the original 
chiral soliton model \cite{DPP97}. This cancellation 
has been argued to be exact in the large $N_c$ limit \cite{Praszalowicz:2003tc}. 
However, a detailed comparison of different versions of the chiral soliton model 
suggests that there is only one coupling constant to leading order and this cancellation 
cannot take place \cite{Walliser:2005pi}. Recently, it has been argued \cite{CoHoLe} 
that heavy pentaquark states $\bar Q q^4$
should be narrow in the combined $1/m_{Q}$ and $1/N_c$ limit. Unfortunately, the 
experimental situation for the heavy pentaquark states $\bar Q q^4$ is also 
inconclusive. The charmed pentaquark initially reported by the H1 Collaboration \cite{posexheavy} 
has not yet been confirmed \cite{nullexheavy,Lee:2005pn}. 

We agree with the conclusions of Refs.~\cite{Walliser:2005pi,Jenkins:2004vb} about 
the existence of a single operator mediating $\Theta \to NK$ transitions at leading order, 
but we find an overall suppression factor of $1/\sqrt{N_c}$. This gives the 
$N_c$ scaling of the strong widths $\Gamma(\Theta \to NK) \sim O(1/N_c)$ for the
positive parity pentaquarks. The corresponding pion widths among different
$\Theta$ states scale like 
$\Gamma(\Theta \to \Theta' \pi) \sim O(N_c^0)$ if the transition $\Theta \to \Theta'$
is allowed by phase space. Any states for which the pion modes are not allowed, which 
include the lowest lying pentaquark, are thus predicted to be narrow 
in the large $N_c$ limit. 

The paper is organized as follows: In Section 2 we construct the complete set of the
positive parity pentaquarks and give their tower structure in the large $N_c$
limit. In Sec.~3 we discuss the strong couplings of the light pentaquarks in the language of
quark operators. Sec.~4 derives the large $N_c$
predictions for the ratios of strong couplings from a consistency condition for 
$\pi + \Theta \to K + B $ scattering.
In Sec.~5 we discuss the heavy pentaquarks in the combined large $N_c$ and heavy quark 
symmetry limit. Finally, Sec.~6 summarizes our conclusions.

\section{Constructing the states}

\begin{figure}[t!]
\begin{center}
  \includegraphics[width=4in]{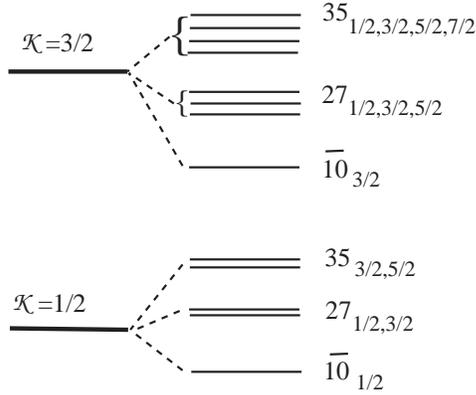}
{\caption{\label{fig1}Schematic representation of the mass spectrum of the 
positive parity pentaquarks. In the flavor symmetric large $N_c$ limit, all
these states are degenerate into two irreducible representations of the
contracted symmetry, labelled with ${\cal K}=1/2, 3/2$. The splittings within
each tower are of order $\sim 1/N_c$. }}
\end{center}
\end{figure}

We start by discussing the construction of the exotic states, using the 
language of the constituent quark model in the large $N_c$ limit.
The quantum numbers of a $\bar q q^{N_c+1}$
hadron are constrained by the fact that the $N_c+1$ quarks have to be in the
fundamental representation of the color $SU(N_c)$ group. Fermi symmetry
requires their spin-flavor-orbital wave function to be in the mixed symmetric
representation ${\cal MS}_{N_c+1} = [N_c,1,0,\cdots]$,  where $[n_1,n_2,\cdots]$ give the
number of boxes in the first, second, etc. row of the corresponding Young tableau. This  spin-flavor-orbital wavefunction can be decomposed into products of 
irreps of $SU(2F)\otimes O(3)$, i.e. spin-flavor wavefunctions  with 
increasingly higher orbital angular momentum
\begin{eqnarray}
{\cal MS}_{N_c+1} \to (MS_{N_c+1}\,,L=0) \oplus (S_{N_c+1}\,, L=1) \oplus \cdots
\end{eqnarray}

The first term corresponds to negative parity exotic states. 
Their properties have been considered in the $1/N_c$ expansion in 
Refs.~\cite{MW,PiSc2}. The second term corresponds to states with a symmetric $SU(2F)$ 
spin-flavor wave function for the $q^{N_c+1}$ system, carrying one unit
of orbital angular momentum $L=1$. After adding in 
the antiquark, this produces positive parity exotics. A subset of these
states were studied using the $1/N_c$ expansion in Ref.~\cite{Jenkins:2004vb}. We reconsider
them here, including all the states dictated by the contracted
$SU(6)_c$ symmetry.

Adopting a Hartree description, one can think of the $q^{N_c+1}$ system
as consisting of $N_c$ quarks in $s$-wave orbitals, plus one
excited quark in a $p$-wave orbital. We write this schematically as
\begin{eqnarray}\label{Hartree}
\Theta =  \bar q q^{N_c}  q^* \ , 
\end{eqnarray}
with $q^*$ denoting the orbitally excited quark. The spin-flavor of the
excited quark is correlated with that of the $ q^{N_c} $ system
such that the total system is in a symmetric representation of $SU(2F)$,
with $F$ the number of light quark flavors. 

For $F=3$ 
the minimal set of these states contains two irreducible representations
of the contracted $SU(6)_{c}$ symmetry, with ${\cal K}=1/2$ and ${\cal K}=3/2$.
The first few states in each of these representations are \cite{PiSc2} (see Fig.~\ref{fig1})
\begin{eqnarray}\label{K12}
{\cal K} =\frac12: && \mathbf{\overline{10}}_\frac12\,,\quad
\mathbf{27}_{\frac12, \frac32}\,,\quad \mathbf{35}_{\frac32, \frac52}\,,\cdots \ , \\
\label{K32}
{\cal K} =\frac32: && \mathbf{\overline{10}}_\frac32\,,\quad
\mathbf{27}_{\frac12, \frac32, \frac52}\,,\quad 
\mathbf{35}_{\frac12, \frac32, \frac52, \frac72}\,,\cdots\,.
\end{eqnarray}
We use the ${\cal K}$ label to denote an entire  $SU(6)_{c}$ representation
by the $SU(4)_c$ multiplets containing the states
with maximal strangeness, sitting at the top of the corresponding
weight diagrams of $SU(3)$.
For the case considered in (\ref{K12}), (\ref{K32}) these are the strangeness ${\cal S}=+1$
states with quark content $\bar s q^{N_c+1}$. We recall here 
that an irreducible  representation of $SU(4)_c$
(tower with fixed strangeness) is labelled by ${\cal K}=0,1/2,1,\dots$ and contains all states
with spin $J$ and isospin $I$ satisfying $|I-J| \leq {\cal K} \leq I+J$.
The first set of ${\cal K}=1/2$ states has been considered in Ref.~\cite{Jenkins:2004vb}. 
The treatment adopted here can
describe both towers.
In this paper we will also consider the ${\cal K}=3/2$ tower in detail.

As the antiquark mass $m_Q$ is increased, the two towers become
closer in mass, split only by effects of order 
$O(1/m_Q)$ as a consequence of heavy quark symmetry \cite{IsWi}. This corresponds to the charmed or bottom exotic states $\bar Q q^{N_c+1}$,
with $Q = c,b$. A more appropriate description for these states is given \cite{PiSc2} in terms of 
one tower for the light degrees of freedom with ${\cal K}_\ell=1$ 
\begin{eqnarray}\label{set1}
{\cal K}_\ell =1: && \mathbf{\overline{6}}_1\,,\quad
\mathbf{15}_{0,1,2}\,,\quad \mathbf{15'}_{1,2,3}\,,\cdots
\end{eqnarray}
where the subscript denotes the spin of the light degrees of freedom $J_\ell$.
Each of these multiplets corresponds to a heavy quark spin doublet, with
hadron spin $J = J_\ell \pm 1/2$, except for the singlets with $J_\ell = 0$.
The properties of
these states are studied below in Sec.~\ref{secheavy}.

\begin{figure}
\begin{center}
\begin{picture}(300,100)(0,0)
\put(0,80){$|[(L,(S_{\bar q}, S_q)S]JI\rangle$}
\put(0,60){quark model states}
\put(200,80){$|[(L,S_{\bar q}){\cal K}, S_q]JI\rangle$}
\put(220,60){tower states}
\put(100,20){$|[(L,S_q)J_\ell,S_{\bar q}]JI\rangle$}
\put(90,00){heavy pentaquarks}
\put(110,80){\vector(1,0){70}}
\put(125,60){Eq.~(\ref{recoupl})}
\put(50,50){\vector(3,-1){50}}
\put(20,35){Eq.~(\ref{recoupl2})}
\put(240,50){\vector(-3,-1){50}}
\end{picture}
{\caption{\label{fig2}The three possible couplings of the angular momenta in a 
pentaquark state with orbital excitation. The connection between the
basis states is given by recoupling relations, given in the text. }}
\end{center}
\end{figure}
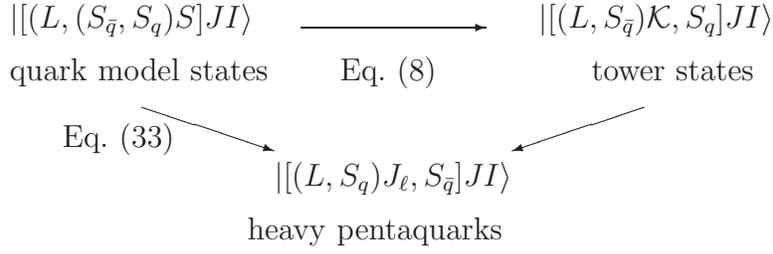

Next we discuss the relation between the different coupling schemes 
when constructing the pentaquark states $ |\Theta; JI \rangle $ in terms of spin and
orbital states. They are obtained by combining the system of $N_c+1$ light quarks
in a spin-flavor symmetric state $|S_q = I\rangle$ with the
orbital angular momentum $|L=1\rangle$ and the antiquark $|S_{\bar q} = \frac12\rangle$
\begin{eqnarray}
|\Theta; JI \rangle \in |q^{N_c+1}; S_q = I\rangle \otimes 
|L=1\rangle \otimes |\bar q; S_{\bar q}=\frac12 \rangle \ .
\end{eqnarray}
The total hadron spin $\vec J$ is thus given by
\begin{eqnarray}
\vec J = \vec S_q + \vec S_{\bar q} + \vec  L \ .
\end{eqnarray}

The three angular momenta $\vec S_q , \vec S_{\bar q}, \vec  L$ can be
coupled in several ways, which give different pentaquark states (see Fig.~\ref{fig2}).
The large $N_c$ QCD states are obtained by coupling these three vectors
in the order $ \vec  L  + \vec S_{\bar q} = \vec {\cal K}$,
$\vec {\cal K} + \vec S_q = \vec J$, with ${\cal K}$ taking the two possible values ${\cal K}=1/2, 3/2$. 
These states will be denoted $|[(L,S_{\bar q}){\cal K}, S_q]JI;m\alpha\rangle$,
with $I=S_q$, and can be identified in the large $N_c$ limit with the
two towers  
corresponding to ${\cal K}=1/2, 3/2$.

Another possible choice for the pentaquark states involves 
coupling first $ \vec S_{\bar q}+ \vec S_q =\vec S$,
with $\vec S$ the total spin of the quark-antiquark system. 
In a second step, the spin $\vec S$ is coupled with the orbital angular momentum $\vec L$
as $\vec L + \vec S  = \vec J$, with $\vec J$ the total spin of the hadron.
We will denote these
states as $|[(L,(S_{\bar q}, S_q)S]JI; m\alpha\rangle$, and they are
the most convenient choice for quark model computations of matrix elements.  Note that 
this coupling scheme is arbitrary since $S$ is not a good quantum number. On the other 
hand ${\cal K}$ is the right quantum number that labels the physical states
in the well defined large $N_c$ limit of QCD.

The connection between the tower states and the quark model states is
given by the usual recoupling formula for 3 angular momenta
\begin{eqnarray}\label{recoupl}
&& |[(L,S_{\bar q}){\cal K}, S_q]JI;m\alpha\rangle
= (-)^{I+1/2+L+J} \\
&& \hspace{2cm} \times \sum_{S=I\pm 1/2} \sqrt{[S] [{\cal K}]}
\left\{
\begin{array}{ccc}
I & \frac12 & S \\
L & J & {\cal K} \\
\end{array}
\right\}
|[(L,(S_{\bar q}, S_q)S]JI; m\alpha\rangle \ , \nonumber
\end{eqnarray} 
where $[S]=2 S+1$, etc. 
This recoupling relation fixes the mixing matrix of the pentaquark states in the
large $N_c$ limit. This is analogous to a result found for the $\mathbf{70}^-$ orbitally 
excited baryons,
for which the
mixing angles are determined in the large $N_c$ limit, up to configuration mixing 
effects \cite{PY1}.

Finally, another possible choice for the pentaquark states combines
first the light quark spin with $\vec L$ into the angular momentum of
the light degrees of freedom $\vec J_\ell = \vec L + \vec S_q $ , which is then coupled
with the antiquark spin to the total spin of the baryon $\vec J = \vec J_\ell + \vec S_{\bar q}$.
The corresponding states will
be denoted as
$|[(L, S_q)J_\ell, S_{\bar q}]JI; m\alpha\rangle $  and are appropriate in the
heavy antiquark limit, when the spin of the light degrees of freedom is
a conserved quantum number. A detailed discussion of these states is
given in Sec.~5.

\section{Light pentaquarks}

We start by discussing the mass spectrum of the positive
parity states. The formalism is very similar to that
used for the $L=2$ baryons in the $\mathbf{56}^+$ \cite{symml2} and $p-$wave
orbitally excited charm baryons \cite{charm}.
The mass operator is a linear combination of the most general 
isoscalar parity even operators constructed from the
orbital angular momentum  $L^i$ and the generators of the $SU(6)_q \otimes SU(6)_{\bar q}$ spin-flavor algebra \cite{Jenkins:2004vb}, the quark operators $S_q, G_q^{ia}, T_q^a$
and  the antiquark operators $S_{\bar q}, G_{\bar q}^{ia}, T_{\bar q}^a$.

For simplicity we restrict our consideration here to the ${\cal S}=+1$ pentaquarks 
with quark content $\bar s q^{N_c+1} $
and assume isospin symmetry.
To leading order in the $1/N_c$ expansion, the mass operator 
acting on these states reads
\begin{eqnarray}
\hat M = c_0 N_c {\mathbf{1}} + c_1 L^i S_{\bar q}^i 
+ O(1/N_c) \ , 
\end{eqnarray}
where $c_0$ and $c_1$ are unknown constants.  
Spin-flavor symmetry is broken at leading order in $1/N_c$ by only 
one operator, describing the spin-orbit interaction of the antiquark.
This operator is diagonal in the ${\cal K}$ basis given by Eq.~(\ref{recoupl})
and is responsible for the $O(N_c^0)$ splitting of the two towers with
${\cal K}=1/2$ and $3/2$. The situation is analogous to the one we have for the non-strange 
members of the $\mathbf{70}^-$ 
excited baryons, where three irreducible representations of the ${\cal K}$-symmetry
have different masses because there are three
operators in the expansion of the mass operator up to $O(N_c^0)$ \cite{PiSc}.

The states at the top of the
weight diagram of a given $SU(3)$ representation (with maximal strangeness)
can decay into a nonstrange ground state baryon and a kaon $\Theta \to BK$.
These transitions are mediated by the strangeness changing axial current and 
can be parameterized in terms of an operator $Y^{i\alpha}$ defined as
\begin{eqnarray}
\langle B| \bar s \gamma^i \gamma_5 q^\alpha |\Theta \rangle =
 (Y^{i\alpha})_{B\Theta}  \ , 
\end{eqnarray}
with $B= N,\Delta, \dots$
The operator $Y^{i\alpha}$ can be expanded in $1/N_c$ as
$Y^{i\alpha} = Y^{i\alpha}_0 + Y^{i\alpha}_1/N_c + \dots$, where the leading 
term scales like $O(N_c^0)$, as will 
be shown below in Sec.~\ref{Ncounting}.

At leading order in $1/N_c$, the explicit representation of $Y^{i\alpha}_0$ in 
the quark operator 
expansion gives only one operator  
\begin{eqnarray}\label{Axdef}
Y^{i\alpha} = g_0  \bar s \xi^i q^{*\alpha} + O(1/N_c) \ , 
\end{eqnarray}
where $\alpha = \pm 1/2$ denotes the flavor of the orbitally excited
light quark $q^*=u,d$ and $g_0$ is an unknown constant that stands for 
the reduced matrix element of the QCD operator. In addition to the 
$SU(4)_q \otimes O(3)$ generators we now 
need as another basic building block  an isoscalar vector
operator acting on the orbital degrees of freedom, which 
we denote $\xi^i$.

In the following we compute the $\Theta \to N,\Delta$
matrix elements of the axial current operator, Eq.~(\ref{Axdef}), and show that they 
can be expressed in terms of a few reduced matrix elements whose expressions 
are already known for arbitrary $N_c$. 
The matrix elements of the operator in Eq.~(\ref{Axdef}) take the simplest form 
when expressed
using the quark model states on the right-hand side of Eq.~(\ref{recoupl}).
They can be computed straightforwardly with the result
\begin{eqnarray}\label{Ax}
&& \langle I'm'\alpha' | \bar s \xi^i q^{*\beta} |[L,(\frac12 I)S]JI;m\alpha
\rangle = \\
&& \qquad  \frac{1}{\sqrt{N_c+1} } 
 \langle 0 |\!| \xi |\!| L\rangle \delta_{L,1}
\delta_{S,I'} \sqrt{\frac{[J]}{[I']}}
t(I',I)
\left(
\begin{array}{cc|c}
J & 1 & I' \\
m & i & m' \\
\end{array}
\right)
\left(
\begin{array}{cc|c}
I & \frac12 & I' \\
\alpha & \beta & \alpha' \\
\end{array}
\right) \ ,
\nonumber
\end{eqnarray} 
where $t(I',I)$ is the reduced matrix element of the ${\bar s} q^\beta$
operator on spin-flavor symmetric states of the $q^{N_c+1}$ system, defined as
\begin{eqnarray}\label{tdef}
\langle I'm'\alpha' |{\bar s} q^\beta | SI;m\alpha\rangle
= t(I',I)\delta_{SI'}\delta_{m'm} 
\left(
\begin{array}{cc|c}
I & \frac12 & I' \\
\alpha & \beta & \alpha' \\
\end{array}
\right) \ .
\end{eqnarray}
The reduced matrix elements $t(I',I)$ for arbitrary $N_c$ can be obtained easily 
using the occupation number formalism as described in Ref.~\cite{Pirjol:2006jp}. 
Explicit results for $t(I',I)$ for all pentaquarks with quantum numbers of interest
are tabulated in Ref.~\cite{Pirjol:2006jp}.
For completeness, we reproduce here the expressions needed in the following.
\begin{eqnarray}\label{4t}
&& t(\frac12,0) = \frac12\sqrt{N_c+1}\,,\hspace{1.6cm}
t(\frac12,1) = \frac{\sqrt3}{2}\sqrt{N_c+5}\,,\\
&& t(\frac32,1) = \frac12 \sqrt{\frac32} \sqrt{N_c-1}\,,\qquad
t(\frac32,2) =\frac12 \sqrt{\frac52} \sqrt{N_c+7} \,.\nonumber
\end{eqnarray}

The explicit suppression factor of $O(1/{\sqrt{N_c})}$ in Eq.~(\ref{Ax}) arises because 
we have to annihilate the excited quark $q^*$ carrying one unit of angular momentum.
The detailed derivation of this factor is given below in Section 4. 
This suppression factor is absent in the case of negative parity pentaquarks 
where the $N_c+1$ quarks are all in $s-$wave orbitals, and the axial current 
can annihilate any of them. The reduced matrix element in Eq.~(\ref{tdef}) 
scales like $t(I',I) \sim N_c^{1/2}$ \cite{Pirjol:2006jp}, which implies that the
$\Theta \to B$ matrix element of the axial current scales like 
$\langle Y^{i\alpha}\rangle \sim N_c^0$. This means that the $\Theta \to BK$
partial widths of these
exotic states are suppressed as $1/N_c$ in the large $N_c$ limit.

Finally, the matrix elements of the axial current on the
tower (physical) pentaquark states are obtained by substituting 
Eq.~(\ref{Ax}) into the recoupling relation
Eq.~(\ref{recoupl}). Because of the
Kronecker symbol $\delta_{S,I'}$, only one of the two
terms on the right-hand side of Eq.~(\ref{recoupl}) gives a nonvanishing 
contribution.
The result is given by
\begin{eqnarray}\label{Kgeneral}
&& \langle I'm'\alpha' | Y^{i\beta}_0 |[(L,\frac12){\cal K}, S_q]JI;m\alpha
\rangle \equiv g_0 T(I',IJ{\cal K}) 
\left(
\begin{array}{cc|c}
J & 1 & I' \\
m & i & m' \\
\end{array}
\right)
\left(
\begin{array}{cc|c}
I & \frac12 & I' \\
\alpha & \beta & \alpha' \\
\end{array}
\right)
\\
&& \qquad
= {\bar g}_0 
t(I',I) \sqrt{[{\cal K}][J]}
\left\{
\begin{array}{ccc}
I & \frac12 & I' \\
1 & J & {\cal K} \\
\end{array}
\right\}
\left(
\begin{array}{cc|c}
J & 1 & I' \\
m & i & m' \\
\end{array}
\right)
\left(
\begin{array}{cc|c}
I & \frac12 & I' \\
\alpha & \beta & \alpha' \\
\end{array}
\right) \ ,
\nonumber
\end{eqnarray}
with ${\bar g}_0$  an overall constant of order $N_c^0$ where we also absorbed the
unknown orbital overlap matrix element.

We list in Tables \ref{K12table} and \ref{K32table} the reduced axial
matrix elements $T(I',IJ{\cal K})$ following from Eq.~(\ref{Kgeneral}), corresponding to 
the two towers with ${\cal K}=1/2$ and ${\cal K}=3/2$, respectively. 
These tables show also the ratios of the
$p-$wave $\Theta \to BK$ partial widths of these states. They are obtained as 
usual, by summing over final states and averaging over initial states
\begin{eqnarray}
\Gamma_{p\rm-wave} &=& g_0^2 
\frac{[I']^2}{[J][I]} 
|T(I',IJ{\cal K})|^2 |\vec p|^3 \ .
\end{eqnarray}
The predictions for the ratios of the decay amplitudes for the
${\cal K}=1/2$ states agree with those of Ref.~\cite{Jenkins:2004vb} 
after taking $N_c=3$. The results for arbitrary $N_c$ and the ratios for the  
  ${\cal K}=3/2$ states are new.

\begin{table}
\caption{\label{K12table} Large $N_c$ predictions for the 
$p$-wave strong decay amplitudes 
of the positive parity light pentaquarks in the ${\cal K}=1/2$ tower.
The last column shows the ratios of the partial $p-$wave rates, normalized to the
$\overline{\mathbf{10}}_{1/2} \to NK$ width, and
with the phase space factor $|\vec p|^3$ removed.}
\begin{center}
\begin{tabular}{rccc}
\hline\hline
Decay  & $T(I',IJ\frac12)$ & $\frac{1}{p^3}\Gamma_{N_c=3}^{\rm ({\it p}-wave)}$ & $\frac{1}{p^3}\Gamma_{N_c \to \infty}^{\rm ({\it p}-wave)}$\\
\hline \hline
$\overline{\mathbf{10}}_{1/2} \to NK$ & $\frac{1}{\sqrt{N_c+1}}t(\frac12, 0)$ 
                                     & $1$ & $1$\\
\hline
${\mathbf{27}}_{1/2} \to NK$ & $\frac{1}{3\sqrt{N_c+1}} t(\frac12, 1)$ & $\frac29$ & $\frac19$\\
$\to \Delta K$ & $- \frac{2}{3\sqrt{N_c+1}} t(\frac32, 1)$ & $\frac49$ & $\frac89$\\
${\mathbf{27}}_{3/2} \to NK$ & $-\frac{2\sqrt2}{3\sqrt{N_c+1}} t(\frac12, 1)$ & $\frac89$ & $\frac49$\\
$\to \Delta K$ & $\frac{\sqrt5}{3\sqrt{N_c+1}} t(\frac32, 1)$ & $\frac5{18}$ & $\frac59$ \\
\hline
${\mathbf{35}}_{3/2} \to \Delta K$ & $\frac{1}{\sqrt{5(N_c+1)}} t(\frac32, 2)$ & $\frac12$ & $\frac15$\\
${\mathbf{35}}_{5/2} \to \Delta K$ & $-\frac{2}{\sqrt{5(N_c+1)}} t(\frac32, 2)$ & $\frac{4}{3}$ & $\frac{8}{15}$\\
\hline
\hline
\end{tabular}
\end{center}
\end{table}

\begin{table}
\caption{\label{K32table} Large $N_c$ predictions for the 
$p$-wave strong decay amplitudes 
of the positive parity light pentaquarks in the ${\cal K}=3/2$ tower.
The last column shows the ratios of the partial $p-$wave rates, normalized to the
$\overline{\mathbf{10}}_{3/2} \to NK$ width, 
with the phase space factor $|\vec p|^3$ removed.}
\begin{center}
\begin{tabular}{rccc}
\hline \hline
Decay & $T(I',IJ\frac32)$ &  $\frac{1}{p^3}\Gamma^{\rm ({\it p}-wave)}_{N_c=3}$ & 
$\frac{1}{p^3}\Gamma_{N_c \to \infty}^{\rm ({\it p}-wave)}$ \\
\hline \hline
$\overline{\mathbf{10}}_{3/2} \to NK$ & $-\frac{\sqrt2}{\sqrt{N_c+1}} t(\frac12, 0)$ & $1$ &  $1$\\
\hline
${\mathbf{27}}_{1/2} \to NK$ & $-\frac{2\sqrt2}{3\sqrt{N_c+1}} t(\frac12, 1)$ & $\frac{16}{9}$ & $\frac{8}{9}$\\
$\to \Delta K$ & $- \frac{1}{3\sqrt{2(N_c+1)}} t(\frac32, 1)$ & $\frac{1}{18}$ & $\frac{1}{9}$\\
${\mathbf{27}}_{3/2} \to NK$ & $\frac{\sqrt{10}}{3\sqrt{N_c+1}} t(\frac12, 1)$ & $\frac{10}{9}$ & $\frac{5}{9}$ \\
$\to \Delta K$ & $\frac{2}{3\sqrt{N_c+1}} t(\frac32, 1)$ & $\frac29$ & $\frac{4}{9}$\\
${\mathbf{27}}_{5/2} \to \Delta K$ & $-\sqrt{\frac{3}{2(N_c+1)}} t(\frac32, 1)$ & $\frac12$ & $1$ \\
\hline
${\mathbf{35}}_{1/2} \to \Delta K$ & $\frac{1}{\sqrt{2(N_c+1)}} t(\frac32, 2)$ & $\frac52$ & $1$\\
${\mathbf{35}}_{3/2} \to \Delta K$ & $-\frac{2}{\sqrt{5(N_c+1)}} t(\frac32, 2)$ & 2 & $\frac{4}{5}$  \\
${\mathbf{35}}_{5/2} \to \Delta K$ & $\frac12\sqrt{\frac{14}{5(N_c+1)}} t(\frac32, 2)$ & $\frac76$ & $\frac{7}{15}$\\
\hline \hline
\end{tabular}
\end{center}
\end{table}

In the large $N_c$ limit the ratios of strong decay widths satisfy sum rules.
These sum rules express the equality of the widths of each tower state in 
each partial wave, and are a consequence of the contracted $SU(4)_c$ symmetry, which
relates all tower states in the large $N_c$ limit. They are given by  
\begin{eqnarray}\label{sr1}
\Gamma(\Theta_{R_J} \rightarrow N K ) + \Gamma(\Theta_{R_J} \rightarrow \Delta K )  
&=& \Gamma(\Theta_{\mathbf{\overline{10}}_{1/2}} \rightarrow N K ) \ ,
\end{eqnarray}
where $R_J = {\mathbf{27}}_{1/2}, {\mathbf{27}}_{3/2}, {\mathbf{27}}_{5/2}, 
{\mathbf{35}}_{1/2}, \dots $. These sum rules can be checked explicitly using the results 
listed in the last column of Tables~\ref{K12table} and~\ref{K32table}. 

These sum rules are not apparent in the results of Ref.~\cite{Jenkins:2004vb}, which correspond to 
the case of finite $N_c=3$, for which the contracted symmetry is broken.

Walliser and Weigel \cite{Walliser:2005pi} discussed the pentaquark strong width in
the chiral soliton model. They 
found that only one operator contributes to the $\Theta \to NK$ coupling at leading
order in the $1/N_c$ expansion and  give the prediction
$\Gamma (\Theta_{\mathbf{27}_{3/2}})/ \Gamma(\Theta_{\mathbf{\overline{10}}_{1/2}}) = 
4/9$, which is in agreement with our model independent result in Table~\ref{K12table}.

\section{Large $N_c$ power counting and consistency conditions}
\label{Ncounting}

The results of the preceding section on the $\Theta \to BK$ strong couplings 
take a particularly simple form at leading order in the $1/N_c$ expansion.
This follows in a model-independent way from a consistency condition
satisfied by the matrix elements of $Y^{i\alpha}$, similar to a consistency
condition constraining kaon couplings to ordinary baryons \cite{DJM}.

We start by deriving in some detail the $Y^{i\alpha} \sim N_c^0$ scaling of the 
leading term,
which follows from the special structure of the exotic states with positive
parity considered in this work.
In particular, we show that taking into account the nonzero orbital angular momentum
$L=1$ of these states is crucial in order to obtain the correct $N_c$ scaling.

As explained in Sec.~2, the exotic state can be written schematically as
 $\Theta  = {\bar q} q^{N_c} q^* $, where the
$q^*$ quark carries one unit of angular momentum. The $N_c+1$ quarks are in a 
completely symmetric spin-flavor wave function and a completely antisymmetric 
color-orbital wave function. This state can be described as a linear combination of
terms with given occupation numbers for one-particle states
\begin{eqnarray}\label{occupno}
|q^{N_c} q^* \rangle = \sum_{\{n_i\}} c_{\{n_i\}} | \{ n_1, n_2, n_3, n_4 \}
\rangle \otimes |[n_s^1, n_s^2, \cdots , n_s^{N_c} ; n_p^1, n_p^2, \cdots , n_p^{N_c} ] \rangle  \ .
\end{eqnarray}
The first factor denotes the occupation numbers of the four spin-flavor one-quark
states $\{ u_\uparrow, u_\downarrow, d_\uparrow, d_\downarrow \}$ as defined in 
Ref.~\cite{Pirjol:2006jp}. The second factor denotes the occupation numbers of the $2N_c$
possible orbital-color one-quark states. These are $s-$ and $p-$wave orbitals,
times the $N_c$ possible color states $\phi_s(x) \otimes |i\rangle$ and $\phi_p(x) \otimes
|i\rangle$, respectively. $\{n_i\}$ denotes the set of all occupation numbers.
We consider in this paper only states with one quark in a $p-$wave orbital, and 
denote the color-orbital wave function of such states as $[ N_c ; 1]$.

The axial current is given by the operator in Eq.~(\ref{Axdef})
\begin{eqnarray}
Y_0^{i\alpha} = g_0 \bar s \xi^i q^{*\alpha} \ ,
\end{eqnarray}
where $q^*$ annihilates the spin-flavor state of 
the orbitally excited quark and $\xi^i$ acts on the orbital
wave function of that  state.
When acting on the state, Eq.~(\ref{occupno}), this operator annihilates one quark in a $p-$wave 
orbital. Taking for definiteness $q^* = u_\uparrow^*$, the matrix element of the axial 
current reduces to evaluating expressions of the form
\begin{eqnarray}\label{20}
 u^{*}_\uparrow |q^{N_c} q^* \rangle &=&
 \sum_{\{n_i\}} c_{\{n_i\}}  u_\uparrow^{*}|\{ n_1, n_2, n_3, n_4 \} \rangle 
\otimes |[N_c ; 1 ]\rangle \ .
\end{eqnarray}
The action of the annihilation operator $u_\uparrow^*$ on the symmetric spin-flavor
state can be computed using the methods of Ref.~\cite{Pirjol:2006jp}. However, one subtle point
is that this operator can only annihilate the excited quark, but not the other $N_c$
$s-$wave quarks.

The spin-flavor state of the excited quark can be made explicit with the help of the
identity
\begin{eqnarray}
&& \{ n_1, n_2, n_3, n_4 \} = 
\sqrt{\frac{n_1}{N_c+1}} (u_\uparrow^{*\dagger}) \{n_1-1, n_2, n_3, n_4 \} +
\sqrt{\frac{n_2}{N_c+1}} (u_\downarrow^{*\dagger}) \{n_1, n_2-1, n_3, n_4 \} \nonumber \\
&& \qquad \qquad +
\sqrt{\frac{n_3}{N_c+1}} (d_\uparrow^{*\dagger}) \{n_1, n_2, n_3-1, n_4 \} +
\sqrt{\frac{n_4}{N_c+1}} (d_\downarrow^{*\dagger}) \{n_1, n_2, n_3, n_4-1 \} \ , 
\end{eqnarray}
where $ n_1+n_2+n_3+n_4 = N_c+1$. Using this identity, Eq.~(\ref{20}) can be
evaluated explicitly with the result
\begin{eqnarray}
 u^{*}_\uparrow |q^{N_c} q^* \rangle &=&
 \sum_{\{n_i\}} c_{\{n_i\}}  \sqrt{\frac{n_1}{N_c+1}}|\{ n_1-1, n_2, n_3, n_4 \} \rangle
\otimes| [N_c ; 0 ]\rangle 
\end{eqnarray}
and similarly for other spin-flavor states of the excited quark. 
These relations generalize the relations given in Ref.~\cite{Pirjol:2006jp} for the action of
one-body operators in the occupation number formalism to the case of more complicated
orbital wave functions. 
Note the additional suppression factor $1/\sqrt{N_c+1}$, which would not be present if
all $N_c+1$ quarks were in $s-$wave orbitals. Together with Eq.~(\ref{Ax}) given in 
the previous section,
this completes the proof of the $N_c$ scaling of the $\Theta \to B$ matrix elements of the
axial current.

This scaling implies that the
decay amplitude $A(\Theta \to BK)$ scales like $N_c^{-1/2}$, which in turn 
predicts that the corresponding strong decay widths are parametrically
suppressed by $1/N_c$. This suppression may be obscured in the total 
widths of the $\Theta$ states by two possible mechanisms. First, the pion modes
$\Theta \to \Theta' \pi$, whenever allowed by phase space,  
have widths of order $O(N_c^0)$. Second, mixing of the exotic states with radially
excited nucleon states, such as the Roper resonance, could enhance the $N_c$ scaling 
of the decay amplitude as $A(\Theta \to \pi N) \sim O(N_c^0)$.
None of these mechanisms applies to the lowest lying pentaquark state(s), for which
the $1/N_c$ expansion offers thus another possible explanation for their
small widths.

Accounting explicitly for the $L=1$ orbital momentum of these states
is crucial for obtaining the $Y^{i\alpha}\sim N_c^0$ scaling. 
This can be contrasted with the approach of Ref.~\cite{Jenkins:2004vb}, where the orbital 
angular momentum is not explicit. Instead, the angular momentum $L=1$ is coupled
with the antiquark spin $S_{\bar q}$ to a fixed value ${\cal K}=
L+S_{\bar q} = 1/2$, and ${\cal K}$ is effectively identified with $S_{\bar q}=\frac12$.
The $\xi^i$ operator acting on the orbital part does not appear in any of the operators
describing physical quantities, such as masses, axial currents, etc.
In this approach,
the axial current operator mediating the $\Theta \to B $ transition is identified with
\cite{Jenkins:2004vb}
\begin{eqnarray}
Y_{({\rm no} \ \xi )}^{i\alpha} = g_0 \bar s \sigma^i q^\alpha + O(1/N_c)
\end{eqnarray}
and its matrix elements scale like $N_c^{1/2}$ \cite{Pirjol:2006jp}. 

We turn next to derive the leading behaviour of $Y^{i\alpha}$ in a hadronic 
language.
The matrix elements of the leading
order piece $Y^{i\alpha}_0$ satisfy a consistency condition from 
$\pi^a + \Theta  \to K^\alpha + B $ scattering and can be obtained in 
a model independent way in the large $N_c$ limit. The pion couplings 
to ordinary baryons and pentaquarks are parametrized by 
\begin{eqnarray}
\langle B'| \bar q \gamma^i \gamma_5 \tau^a q |B \rangle &=&
 N_c (X^{ia})_{B'B} , \\ 
\langle \Theta'| \bar q \gamma^i \gamma_5 \tau^a q |\Theta \rangle &=&
 N_c (Z^{ia})_{\Theta'\Theta}  .
\end{eqnarray}
These operators have a $1/N_c$ expansion of the form 
$X^{ia} = X_0^{ia} + X_1^{ia}/N_c + ... $ , 
and similarly for $Z^{ia}$, where the leading order terms $X_0^{ia}$ 
and $Z_0^{ia}$ scale as $O(N_c^0)$.

After including the meson decay constants, the overall scaling of the direct and crossed
diagrams separately 
is $O(N_c^0)$. 
The calculation of the scattering amplitude at the quark level gives a
$1/\sqrt{N_c}$ scaling for the $\pi^a + \Theta \to  K^\alpha + B $ amplitude.
This leads to the consistency condition 
\begin{eqnarray}\label{cc}
Y_0^{j\alpha} Z_0^{ia} - X_0^{ia} Y_0^{j\alpha} = 0 \ . 
\end{eqnarray} 
The derivation is similar to the one given for the consistency condition of 
meson couplings to ordinary and hybrid baryons
in Ref.~\cite{CPY99}.

The leading order matrix element $\langle X^{ia}_{0} \rangle$ is given by the model-independent
expression \cite{Largenspinflavor,DJM}
\begin{eqnarray}
 \langle X_0^{ia}\rangle = g_X
(-)^{J+I'+{\cal K}+1} \sqrt{[I][J]}
\left\{
\begin{array}{ccc}
I' & 1 & I \\
J &  {\cal K} & J' \\
\end{array}
\right\}
\left(
\begin{array}{cc|c}
J & 1 & J' \\
J_3 & i & J'_3 \\
\end{array}
\right)
\left(
\begin{array}{cc|c}
I& 1 & I' \\
I_3 & a & I'_3 \\
\end{array}
\right)
\end{eqnarray}
and similarly for $\langle Z^{ia}_0 \rangle$ with $g_Z$ instead of $g_X$. 
The $\Theta  \to K^\alpha + B $ vertex is parametrized by
\begin{eqnarray}\label{Ksol}
\langle I'I'_3; J'J'_3; {\cal K}' |Y_0^{i\alpha}|II_3; JJ_3; {\cal K} \rangle &=& 
\sqrt{[I][J]} {\cal Y}_0(I'J'{\cal K}';IJ{\cal K}) 
\left(
\begin{array}{cc|c}
J & 1 & J' \\
J_3 & i & J'_3 \\
\end{array}
\right)
\left(
\begin{array}{cc|c}
I & \frac12 & I' \\
I_3 & \alpha & I'_3 \\
\end{array}
\right) \ .
\end{eqnarray}
For ${\cal K}'=0$ 
this expression is equivalent to Eq.(\ref{Kgeneral}), with the identification
$g_0 T(I',IJ{\cal K}) = \sqrt{[I][J]} {\cal Y}_0(I'I'0;IJ{\cal K})$.

Taking the matrix elements of Eq.(\ref{cc}) between $B(I'J'{\cal K}')$  
and $\Theta(IJ{\cal K})$,
and projecting onto channels with total spin $H$ and isospin $T$ in the $s$-channel 
we obtain 
\begin{eqnarray}
&& \sum_{{\bar I},{\bar J}} (-)^{-{\bar I}-\frac{1}{2}} [{\bar I}][{\bar J}] 
\left\{
\begin{array}{ccc}
1 & I'        & {\bar I} \\
{\cal K}' & {\bar J} & J'
\end{array}
\right\}
\left\{
\begin{array}{ccc}
1  & {\bar J}  & J' \\
1  & H         & J
\end{array}
\right\}
\left\{
\begin{array}{ccc}
\frac{1}{2} & {\bar I} & I \\
1 & T & I' 
\end{array}
\right\}
{\cal Y}_0({\bar I}{\bar J}{\cal K}';IJ{\cal K})
= \nonumber \\
&& \qquad = 
(-)^{H+{\cal K}+{\cal K}'-I'-J} \delta(J'1H) \delta(I'\frac12 T) \frac{g_Z}{g_X} 
\left\{
\begin{array}{ccc}
1 & T & I \\
{\cal K} & J & H
\end{array}
\right\}
{\cal Y}_0(I'J'{\cal K}';TH{\cal K}) \ ,
\end{eqnarray}
where $\delta(J'1H)=1$ if $|J'-1|\le H \le J'+1$ and zero otherwise, etc. 
The most general solution of this equation implies $g_X=g_Z$, and depends on two arbitrary
couplings $c_y$ with $y=1/2, 3/2$ 
\begin{eqnarray} \label{y0sol}
{\cal Y}_0(I'J'{\cal K}';IJ{\cal K}) &=&
\sum_{y=1/2,3/2} c_y 
\left\{
\begin{array}{ccc}
\frac12 & 1 & y \\
I & J & {\cal K} \\
I' & J' &  {\cal K}' \\
\end{array}
\right\} \ ,
\end{eqnarray}
up to an arbitrary phase $(-1)^{2nI+2mJ}$ with $n,m$ integers. For decays to
nonstrange baryons ${\cal K}'=0$, and this equation gives the asymptotic form 
for $T(I',IJ{\cal K})$ in the large $N_c$ limit
\begin{equation}
\lim_{N_c \to \infty} T(I',IJ{\cal K}) 
\propto (-)^{1+I+I'+{\cal K}}
\sqrt{\frac{[I][J]}{[I'][{\cal K}]}}  
\left\{
\begin{array}{ccc}
\frac12 & 1 &  {\cal K} \\
J & I  & I' \\
\end{array}
\right\} \ .
\end{equation}
This agrees with the large $N_c$ limit of the
reduced matrix element obtained by the quark operator calculation in Sec.~3.

\section{Heavy pentaquarks}
\label{secheavy}

Taking the antiquark 
to be a heavy quark $Q=c,b$, the quantum numbers of the $\bar Q q^{N_c+1}$
states are simply related to those of the $N_c+1$ quarks, as was discussed earlier in Sec.~2.
These states belong to one 
large $N_c$ tower with ${\cal K}_\ell=1$ and are shown in Eq.~(\ref{set1}).

We pause here to compare these states  with 
the positive parity heavy pentaquarks
considered in \cite{Jenkins:2004vb}. The light quarks in those states belong 
to a ${\cal K}_\ell=0$ tower and 
include the $SU(3)$ representations
\begin{eqnarray}
\label{set0}
{\cal K}_{\ell}=0 & &:\quad \mathbf{\overline{6}}_0\,, \mathbf{15}_{1}\,, 
\mathbf{15'}_{2}\,, \cdots
\end{eqnarray}
where the subscript denotes the spin of the light degrees of freedom $J_\ell$.
Each of these multiplets corresponds to a heavy quark spin doublet, with
hadron spin $J = J_\ell \pm 1/2$, except for the singlets with $J_\ell = 0$.
Note that they are different from the states constructed here in 
Eq.~(\ref{set1}), which in addition to the more complex mass spectrum 
also have very different strong couplings, as will be seen below.

The heavy pentaquark states require a different recoupling of the three
angular momenta $S_q, S_{\bar q}, L$. Neither $S_q$ nor $L$ are good
quantum numbers for a heavy pentaquark, but only their sum, the 
angular momentum of the light degrees of freedom $J_\ell =
S_q + L$, is. The states
with good $J_\ell$ are expressed in terms of the quark model states
by a recoupling relation analogous to Eq.~(\ref{recoupl})
\begin{eqnarray}\label{recoupl2}
&& |[(L,J_{q})J_\ell, S_{\bar q}]JI;m\alpha\rangle
= (-)^{I+1/2+L+J} \\
&& \hspace{2cm} \times \sum_{S=I\pm 1/2} \sqrt{[S][J_\ell]}
\left\{
\begin{array}{ccc}
I & \frac12 & S \\
J & L & J_\ell \\
\end{array}
\right\}
|[(L,(J_{q}, S_{\bar q})S]JI; m\alpha\rangle \ . \nonumber 
\end{eqnarray}
A similar recoupling relation can be written which expresses the
heavy pentaquark states $|[(L,J_{q})J_\ell, S_{\bar q}]JI;m\alpha\rangle$
in terms of tower states $|[(L,S_{\bar q}){\cal K}, J_{q}]JI;m\alpha\rangle$, 
appropriate for the light pentaquark states. Such a relation makes
explicit the correspondence between light and heavy pentaquarks,
as the mass of ${\bar Q}$ is gradually increased. We do not write this
relation explicitly, but just mention one of its implications: any given 
heavy pentaquark state is related to states in both ${\cal K}=1/2$ and $3/2$ towers.
Any treatment which neglects one of these towers will therefore
be difficult to reconcile with a quark model picture.

A generic positive parity heavy pentaquark state $\Theta_{\bar Q}$ can decay 
strongly into the four channels  
\begin{eqnarray}\label{NHQ}
&& \Theta_{\bar Q} \to N H_{\bar Q}, N H_{\bar Q}^* \ ,  \\
\label{DelHQ}
&& \Theta_{\bar Q} \to \Delta H_{\bar Q}, \Delta H_{\bar Q}^*  \ , 
\end{eqnarray}
where $H_{\bar Q}$ is a pseudoscalar $J^P = 0^-$ heavy
meson with quark content $\bar Q q$, and $H_{\bar Q}^*$ is its heavy quark
spin partner with $J^P = 1^-$. Heavy quark symmetry gives relations among
the amplitudes for these decays \cite{IsWi}. However, these relations alone
are in general not sufficient to predict the ratios of the decay widths of the two
modes in Eq.~(\ref{NHQ}), and of the two modes in Eq.~(\ref{DelHQ}). 
On the other hand, large $N_c$ relates the $NH_{\bar Q}$ and $\Delta H_{\bar Q}$
modes. We will show that in the {\em combined} large $N_c$ and heavy 
quark limits it is possible to make also predictions for the ratios of the
$N H_{\bar Q}$ and $N H^*_{\bar Q}$ modes.

We start by considering first the large $N_c$
predictions for the amplitude ratios into $N H_{\bar Q}$ and $\Delta H_{\bar Q}$.
These decays are mediated by the heavy-light axial current, which at leading
order in $1/N_c$ is given by a single operator in the quark operator expansion,
analogous to that mediating kaon decay
\begin{eqnarray}
\langle B| \bar Q \gamma^i \gamma_5 q^\beta |\Theta_{\bar Q} \rangle =
 g_Q ( \bar Q \xi^i q^\beta)_{B\Theta} + O(1/N_c)   \ , 
\end{eqnarray}
with $B= N,\Delta, \dots$ and $g_Q$ an unknown constant.

The matrix elements of this operator can be parameterized in terms of a reduced matrix 
element, defined as
\begin{eqnarray}
\langle I'm'\alpha' | \bar Q \xi^i q^\beta |[(L,S_q)J_\ell, \frac12 ]JI;m\alpha
\rangle = T(I', IJJ_\ell)
\left(
\begin{array}{cc|c}
J & 1 & I' \\
m & i & m' \\
\end{array}
\right)
\left(
\begin{array}{cc|c}
I & \frac12 & I' \\
\alpha & \beta & \alpha' \\
\end{array}
\right) \ .
\end{eqnarray}
At leading order in $1/N_c$, the amplitudes $T(I', IJJ_\ell)$
can be expressed in terms of the  amplitudes $t(I',JI)$ given in Eq.~(\ref{4t}).
To show this, recall that the matrix elements of the axial current 
for the transitions between the
quark model states and the ground state baryons were given in
Eq.~(\ref{Ax}). The corresponding matrix elements taken on the physical heavy pentaquark
states are found by 
substituting this result into the recoupling relation Eq.~(\ref{recoupl2}).
We find
\begin{eqnarray}\label{KgeneralQ}
&& \langle I'm'\alpha' | g_Q \bar Q \xi^i q^\beta |[(L,S_q)J_\ell, \frac12 ]JI;m\alpha
\rangle = \\
&& \qquad
\bar g_Q
\frac{1}{\sqrt{N_c+1}}t(I',I) \sqrt{[J_\ell][J]}
\left\{
\begin{array}{ccc}
I & \frac12 & I' \\
J & 1 & J_\ell \\
\end{array}
\right\}
\left(
\begin{array}{cc|c}
J & 1 & I' \\
m & i & m' \\
\end{array}
\right)
\left(
\begin{array}{cc|c}
I & \frac12 & I' \\
\alpha & \beta & \alpha' \\
\end{array}
\right)  \ , \nonumber
\end{eqnarray}
where the unknown orbital overlap matrix element has also been absorbed in the 
 order $N_c^0$ unknown constant $\bar g_Q$.

We summarize in Table~\ref{h5q}  the 
reduced matrix elements $T(I', IJJ_\ell)$ of the leading order operator in the expansion 
of the heavy-light axial current, for
heavy pentaquarks with positive parity. We denote the pentaquark states as 
$\Theta_{{\bar Q}J_\ell}^{(R)}(J)$,
with $R=\overline{\mathbf 6}, {\mathbf {15}}, {\mathbf {15}'}$ the $SU(3)$ representation
to which they belong. These results give predictions for the ratios of the partial widths of 
$p$-wave strong decays $\Theta_{{\bar Q}J_\ell}^{(R)}(J) \to N H_{\bar Q}, \Delta H_{\bar Q}$.

In the case of heavy pentaquarks there is a second sum rule 
\begin{eqnarray}\label{sr2}
\sum_{J_\ell}\Gamma(\Theta_{{\bar Q} J_\ell}^R(J) \rightarrow N H_{\bar Q})  
&=& \Gamma(\Theta_{{\bar Q} 1}^{\overline{\mathbf 6}}(J) \rightarrow N H_{\bar Q}) \ , 
\end{eqnarray}
in addition to the one
already discussed in Eq.~(\ref{sr1}).
Both hold in the large $N_c$ limit and can be checked explicitly using the results in 
the last column of Table~\ref{h5q}.

\begin{table}
\caption{\label{h5q} Large $N_c$ predictions for the 
$p$-wave heavy pentaquark decay amplitudes
$\Theta_{{\bar Q}J_\ell}^{(R)} \to N H_{\bar Q}$ and 
$\Theta_{{\bar Q}J_\ell}^{(R)} \to \Delta  H_{\bar Q}$. 
In the last column we show the ratios of the partial $p$-wave rates, 
with the phase space factor $p^3$ removed.}
\begin{center}
\begin{tabular}{rccc}
\hline \hline
Decay & $T(I',IJJ_\ell)$ &  $\frac{1}{p^3}\Gamma^{\rm ({\it p}-wave)}_{N_c=3}$ & $\frac{1}{p^3}\Gamma_{N_c \to \infty}^{\rm ({\it p}-wave)}$\\
\hline \hline
$\Theta_ {\bar Q 1}^{(\overline{6})}(\frac12) \to N H_{\bar Q}$ & $ \frac{1}{\sqrt{N_c+1}} t(\frac12,0)$  & 1 & 1  \\
$\Theta_ {\bar Q 1}^{(\overline{6})}(\frac32) \to N H_{\bar Q}$ & $-\sqrt{\frac{2}{N_c+1}} t(\frac12,0)$  & 1 & 1 \\
\hline
$\Theta_ {\bar Q 0}^{({15})}(\frac12) \to N H_{\bar Q}$ & $ \frac{1}{\sqrt{3(N_c+1)}}t(\frac12,1)$ & $\frac23$ & $\frac13$ \\
                        $\to \Delta  H_{\bar Q}$ & $-\frac{1}{\sqrt{3(N_c+1)}}t(\frac32,1)$ & $\frac13$ & $\frac23$ \\
$\Theta_ {\bar Q 1}^{({15})}(\frac12) \to N H_{\bar Q}$ & $-\sqrt{\frac{2}{3(N_c+1)}}t(\frac12,1)$  & $\frac43$ & $\frac23$  \\
                        $\to \Delta  H_{\bar Q}$ & $-\frac{1}{\sqrt{6(N_c+1)}}t(\frac32,1)$  & $\frac16$ & $\frac13$  \\
$\Theta_ {\bar Q 1}^{({15})}(\frac32) \to N H_{\bar Q}$ & $-\frac{1}{\sqrt{3(N_c+1)}}t(\frac12,1)$  & $\frac13$ & $\frac16$  \\
                        $\to \Delta  H_{\bar Q}$ & $ \sqrt{\frac{5}{6(N_c+1)}}t(\frac32,1)$  & $\frac{5}{12}$ & $\frac56$ \\
$\Theta_ {\bar Q 2}^{({15})}(\frac32) \to N H_{\bar Q}$ & $ \sqrt{\frac{5}{3(N_c+1)}}t(\frac12,1)$  & $\frac{5}{3}$ & $\frac56$ \\
                        $\to \Delta  H_{\bar Q}$ & $ \frac{1}{\sqrt{6(N_c+1)}}t(\frac32,1)$  & $\frac{1}{12}$ & $\frac16$ \\
 $\Theta_ {\bar Q 2}^{({15})}(\frac52) \to \Delta  H_{\bar Q}$ & $-\sqrt{\frac{3}{2(N_c+1)}} t(\frac32,1)$  & $\frac12$ & $1$  \\
\hline
$\Theta_ {\bar Q 1}^{({15'})}(\frac12)  \to \Delta  H_{\bar Q}$ & $\frac{1}{\sqrt{2(N_c+1)}} t(\frac32,2)$ & $\frac52$ & $1$ \\
$\Theta_ {\bar Q 1}^{({15'})}(\frac32) \to \Delta  H_{\bar Q}$ & $\frac{1}{\sqrt{10(N_c+1)}} t(\frac32,2)$  & $\frac14$ & $\frac1{10}$ \\
$ \Theta_ {\bar Q 2}^{({15'})}(\frac32)  \to \Delta  H_{\bar Q}$ & 
             $-\frac{3}{\sqrt{10(N_c+1)}} t(\frac32,2)$ & $\frac94$ & $\frac9{10}$\\
$\Theta_ {\bar Q 2}^{({15'})}(\frac52) \to \Delta  H_{\bar Q}$ & 
             $-\frac{1}{\sqrt{10(N_c+1)}} t(\frac32,2)$ & $\frac16$ & $\frac1{15}$\\
$\Theta_ {\bar Q 3}^{({15'})}(\frac52) \to \Delta  H_{\bar Q}$ & 
             $\sqrt{\frac{7}{5(N_c+1)}} t(\frac32,2)$ & $\frac73$ & $\frac{14}{15}$\\
\hline \hline
\end{tabular}
\end{center}
\end{table}

Finally, we will also use heavy quark symmetry to make predictions for modes 
containing a $H_{\bar Q}^*$ heavy meson in the final state. To that end, 
we first describe the
heavy quark symmetry relations and then we combine them with the large $N_c$
predictions. For simplicity, we restrict ourselves to the $N H_{\bar Q}^{(*)}$ modes
in a $p$-wave.

The specification of the final state in 
$\Theta_{{\bar Q}J_\ell} \to [N H^*_{\bar Q}]_{p\rm -wave}$
must include in addition to the total angular momentum ${\vec J} = {\vec S}_N +
{\vec J}_{H^*_Q} + {\vec L}$, also the partial sum of two of the three angular 
momenta. We denote here with ${\vec S}_N$ the spin of the nucleon, ${\vec J}_{H^*_Q}$
the spin of the vector heavy meson, and $ {\vec L}$ the orbital angular momentum.
The heavy quark symmetry relations take a simple form when this partial sum is chosen as
${\vec J}_N = {\vec S}_N + {\vec L}$. The decay amplitude for 
$\Theta_{\bar Q}(IJ J_\ell) \to [N H^{(*)}_{\bar Q}(J'J'_\ell)]_{J_N}$
is given by \cite{IsWi}
\begin{eqnarray}
A(\Theta_{\bar Q}(IJJ_\ell) \to [N H^{(*)}_{\bar Q}(J'J'_\ell)]_{J_N}) = 
\sqrt{[J_\ell][J]}
\left\{
\begin{array}{ccc}
J_\ell & J'_\ell & J_N \\
J' & J & \frac12 \\
\end{array}
\right\} F_{J_\ell J'_\ell J_N}^I \ , 
\end{eqnarray}
where we denoted as usual with $J_\ell$ and  $J'_\ell = 1/2$ the spins of 
the light degrees of
freedom in the initial and final heavy hadrons. $F_{J_\ell J'_\ell J_N}^I$
are reduced matrix elements, which in general also depend on $S_N$, although this
dependence was omitted for simplicity.

The predictions from these relations are shown in explicit
form in Table~\ref{HQStable}, from which one can read off the number of independent
hadronic amplitudes parameterizing each mode. The $\Theta_ {\bar Q 0}(\frac12)$ decays are parameterized
in terms of one reduced amplitude $f_0$, the decays of the $\Theta_ {\bar Q 1}(\frac12,\frac32)$
depend on two independent amplitudes $f_{1,2}$, and the decays of the $\Theta_ {\bar Q 2}(\frac32,\frac52)$
contain one independent amplitude $f_3$. From this counting it follows that heavy quark symmetry
does not relate, in general, all modes with pseudoscalar and vector heavy mesons.

\begin{table}
\begin{center}
\caption{\label{HQStable} Heavy quark symmetry predictions 
for the reduced decay amplitudes  
$\Theta_{{\bar Q}J_\ell} \to [N  H_{\bar Q}^{(*)}]_{p\rm -wave}$ for final states
with given $\vec J_N = \vec S_N + \vec L$.
The zero entries denote
amplitudes forbidden in the heavy quark limit and suppressed
as $\sim O(\Lambda_{QCD}/m_Q)$. }
\begin{tabular}{ccc|ccc}
\hline \hline
Decay & $J_N=1/2$ & $J_N = 3/2$ & Decay & $J_N=1/2$ & $J_N = 3/2$ \\
\hline \hline
$\Theta_{{\bar Q}0}(\frac12) \to NH_{\bar Q}$ & $-\frac{1}{\sqrt{2}} f_0$ & $-$ &
$\Theta_{{\bar Q}0}(\frac12)\to NH^*_{\bar Q}$ & $ \frac{1}{\sqrt2}f_0$ & $0$ \\
\hline
$\Theta_{{\bar Q}1}(\frac12) \to NH_{\bar Q}$ & $\sqrt{\frac{3}{2}} f_1$ & $-$ &
$\Theta_{{\bar Q}1}(\frac12)\to NH^*_{\bar Q}$ & $\frac{1}{\sqrt6}f_1$ & $- \sqrt{\frac{2}{3}}f_2$ \\
$\Theta_{{\bar Q}1}(\frac32) \to NH_{\bar Q}$ & $-$ & $- \sqrt{\frac32} f_2$ &
$\Theta_{{\bar Q}1}(\frac32)\to NH^*_{\bar Q}$ & $- \frac{2}{\sqrt{3}} f_1$ & $\sqrt{\frac56}f_2$ \\
\hline
$\Theta_{{\bar Q}2}(\frac32) \to NH_{\bar Q}$ & $-$ & $\sqrt{\frac52} f_3$ & 
$\Theta_{{\bar Q}2}(\frac32)\to NH^*_{\bar Q}$ & $0$ & ${\frac{1}{\sqrt2}}f_3$ \\
$\Theta_{{\bar Q}2}(\frac52) \to NH_{\bar Q}$ & $-$ & $-$ &
$\Theta_{{\bar Q}2}(\frac52)\to NH^*_{\bar Q}$ & $-$ & $-\sqrt{2}f_3$ \\
\hline \hline
\end{tabular}
\end{center}
\end{table}

Such a relation becomes possible however in the large $N_c$ limit, where all modes with 
a pseudoscalar heavy meson in the final state $N H_{\bar Q}$ are related. In the language of 
the reduced amplitudes in Table
\ref{HQStable}, this amounts to a relation among the amplitudes $f_{0-3}^I$. These
predictions can be obtained by comparing the amplitudes 
listed in Tables 3 and 4. 
We find
\begin{eqnarray}
&& f_0^{I=1} \equiv F^{I=1}_{0\frac12\frac12} = -\sqrt{\frac{N_c+5}{N_c+1}}   \ ,          \\
&& f_1^{I=0} \equiv F^{I=0}_{1\frac12\frac12} = {\frac{1}{\sqrt3}}  \,, \quad
   f_2^{I=0} \equiv F^{I=0}_{1\frac12\frac32} = \sqrt{\frac{2}{3}}  \ ,           \\
&& f_1^{I=1} \equiv F^{I=1}_{1\frac12\frac12} = -\sqrt{\frac{2}{3}} \sqrt{\frac{N_c+5}{N_c+1}}     \,, \quad
   f_2^{I=1} \equiv F^{I=1}_{1\frac12\frac32} = \frac{1}{\sqrt3} \sqrt{\frac{N_c+5}{N_c+1}} \ ,      \\
&& f_3^{I=1} \equiv F^{I=1}_{2\frac12\frac32} = \sqrt{\frac{N_c+5}{N_c+1}}  \ .
\end{eqnarray}

The corresponding predictions for the partial decay rates are shown in Table~\ref{h5qvsJM}.
For comparison, we also show in this table 
the results found in Ref.~\cite{Jenkins:2004vb} for the ${\cal K}_\ell=0$ states.

\begin{table}
\begin{center}
\caption{\label{h5qvsJM} Combined large $N_c$ and heavy quark symmetry predictions 
for the ratios of strong decay rates for heavy pentaquark decays 
$\Theta_{\bar Q} \to N  H_{\bar Q}^{(*)}$. 
In the last line we show for comparison 
also the corresponding predictions for the pentaquark states with positive parity
considered in Ref.~\cite{Jenkins:2004vb}. }
\begin{center} 
\begin{tabular}{ccc}
\hline \hline
& \\[-0.12in]
{$I=0$} & $\Theta_{\bar Q} (\frac12) \to (N H_{\bar Q}) : (N H_{\bar Q}^*)$  
& $\Theta_{\bar Q} (\frac32)\to (N H_{\bar Q}) : (N H_{\bar Q}^*) $ \\[0.08in]
\hline \hline
& \\[-0.12in]
${\cal K}_{\ell}=1$   &  $\frac{1}{2} : \frac{3}{2}  \ \ \  (J_\ell = 1)$ 
                &  $\frac{1}{2} : \frac{3}{2}  \ \ \  (J_\ell = 1)$  \\[0.08in]
\hline
& \\[-0.12in]
${\cal K}_{\ell}=0$      & 1 : 3  \hspace{0.1cm} $(J_\ell = 0)$ & - \\[0.08in]
\hline \hline
\end{tabular}
\end{center}

\vspace{.8cm}

\begin{tabular}{ccc}
\hline \hline
& & \\[-0.12in]
 {$I=1$} & $\Theta_{\bar Q} (\frac12)\to (N H_{\bar Q}) : (N H_{\bar Q}^*)$ 
  & $\Theta_{\bar Q} (\frac32)\to (N H_{\bar Q}) : (N H_{\bar Q}^*) $ \\[0.08in]
\hline \hline
& & \\[-0.12in]
${\cal K}_{\ell}=1$  &  $\frac16     : \frac12    \ \ \  (J_\ell = 0)$  
               &  $\frac1{12}  : \frac7{12} \ \ \  (J_\ell = 1)$     \\[0.08in]
               &  $\frac13     : \frac13    \ \ \  (J_\ell = 1)$  
               &  $\frac5{12}  : \frac14    \ \ \  (J_\ell = 2)$     \\[0.08in]
\hline
& & \\[-0.12in]
${\cal K}_{\ell}=0 $    
           & $1$ : $11$ \hspace{0.1cm} $(J_\ell = 1)$ 
           & $4$ : $8$ \hspace{0.1cm} $(J_\ell = 1)$ \\[0.08in]
\hline \hline
\end{tabular}
\end{center}
\end{table}

\newpage

\section{Conclusions}

In this paper we studied the complete set of light and heavy pentaquark 
states with positive parity at leading order in the $1/N_c$ expansion. 
We discussed the structure of the mass spectrum and the strong decays 
of these states. Both are strongly constrained by the contracted 
spin-flavor $SU(4)_c$ symmetry emerging in the large $N_c$ limit, 
leading to mass degeneracies and sum rules for their decay widths.

The exotic 
states which are 
composed of only light quarks $(\bar s q^{N_c+1})$
belong to two irreducible representations (towers) of this symmetry, with
${\cal K}=\frac{1}{2}$ (containing a $J^P=\frac12^+$ isosinglet), and 
${\cal K}=\frac{3}{2}$ (containing a $J^P=\frac32^+$ isosinglet), respectively. 
The strong transitions between any members of a pair of towers are related by the 
contracted symmetry. The states with ${\cal K}=\frac{1}{2}$ are identical to those  
considered in the large $N_c$ expansion in Ref.~\cite{Jenkins:2004vb}, and we 
find complete agreement 
with their $N_c=3$  predictions for these states. The more general results for arbitrary $N_c$ 
and the ${\cal K}=\frac{3}{2}$ tower
states are new.

Taking the antiquark to be a heavy quark, the two irreducible representations
with ${\cal K}=\frac12,\frac32$ are split only by $O(1/m_{Q})$ hyperfine interactions.
In the heavy quark limit they become degenerate and the spin of the light 
degrees of freedom is a conserved quantum number. When this is combined with the 
large $N_c$ limit a new good quantum number emerges: ${\cal K}_{\ell}$, the tower 
label for the light quarks.

We find that the heavy pentaquarks with positive parity belong to one tower with 
${\cal K}_\ell = 1$.
These are different from the heavy exotic states considered in 
Ref.~\cite{Jenkins:2004vb}, which belong to ${\cal K}_\ell = 0$.
Both sets of states are legitimate from the point of view of the large $N_c$ symmetry, 
although explicit realizations of these states are more natural in different models:
the ${\cal K}_\ell = 0$ states are obtained in a Skyrme model picture, while the
${\cal K}_\ell = 1$ states considered here appear naturally in the constituent quark
model picture. The predictions for the strong decays of the two sets of states differ, as 
shown in Table~\ref{h5qvsJM}, and can be used to discriminate between them.

There is an important difference between our treatment of the transition operators
and that given in Ref.~\cite{Jenkins:2004vb}, due to the fact that we
keep the orbital angular momentum explicit. 
In Sec.~\ref{Ncounting} we show, by explicit computation of the $\Theta \to B$
axial current matrix element in the quark model with $N_c$ colors, that
the strong width of the lowest-lying positive parity exotic state scales like $1/N_c$. 
This provides a natural explanation for the existence of narrow exotic states
in the large $N_c$ limit.

\vspace{0.3cm}\noindent
{\em Acknowledgments:}
The work of D.P. was supported by the DOE under cooperative research 
agreement DOE-FC02-94ER40818 and by the NSF under grant PHY-9970781.
The work of C.S. was supported in part by Fundaci\'on Antorchas, Argentina 
and Fundaci\'on S\'eneca, Murcia, Spain.

\end{document}